\documentclass[twocolumn,
superscriptaddress,
showpacs,%preprintnumbers,
 amsmath,amssymb,
 aps,
 prl,
floatfix,
]{revtex4}

\usepackage{graphicx}% Include figure files
\usepackage{dcolumn}% Align table columns on decimal point
\usepackage{bm}% bold math
\usepackage[T1]{fontenc}
\usepackage{lmodern}
\usepackage{color}
\usepackage{soul}

\newcommand{\mum}{\mu {\rm m}}
\begin{document}
\newcommand{\hlrose}[1]{\hl{#1}}
\sethlcolor{yellow}

\preprint{APS/123-QED}

\title{Penrose-Onsager Criterion Validation \\ in a One-Dimensional Polariton Condensate}% Force line breaks with \\
%\thanks{A footnote to the article title}%

\author{F.~Manni}%
 \email{francesco.manni@epfl.ch}
 \affiliation{%
  Institute of Condensed Matter Physics, \'{E}cole Polytechnique F\'ed\'erale de Lausanne (EPFL), CH-1015 Lausanne, Switzerland}%

\author{K.~G.~Lagoudakis}%
\affiliation{%
  Institute of Condensed Matter Physics, \'{E}cole Polytechnique F\'ed\'erale de Lausanne (EPFL), CH-1015 Lausanne, Switzerland}%
  \affiliation{%
  Current address: E. L. Ginzton Laboratory, Stanford University, Stanford, CA93405-4088, US}%

\author{R.~Andr\'e}%
\affiliation{%
  Institut N\'eel, CNRS, 25 Avenue des Martyrs, 38042 Grenoble, France}%

\author{M.~Wouters}%
\affiliation{TQC, Departement Fysica, Universiteit Antwerpen, Belgium}

\author{B.~Deveaud}%
\affiliation{%
  Institute of Condensed Matter Physics, \'{E}cole Polytechnique F\'ed\'erale de Lausanne (EPFL), CH-1015 Lausanne, Switzerland}%

\date{\today}% It is always \today, today,
             %  but any date may be explicitly specified

\begin{abstract}
We perform quantum tomography on one-dimensional polariton condensates, spontaneously occurring in linear disorder valleys in a CdTe planar microcavity sample. By the use of optical interferometric techniques, we determine the first-order coherence function and the amplitude and phase of the order parameter of the condensate, providing a full reconstruction of the single particle density matrix for the polariton system. The experimental data are used as input to theoretically test the consistency of Penrose-Onsager criterion for Bose-Einstein condensation in the framework of nonequilibrium polariton condensates. The results confirm the pertinence and validity of the criterion for a non equilibrium condensed gas.
\end{abstract}

% PACS, the Physics and Astronomy
                             % Classification Scheme.
%\keywords{Suggested keywords}%Use showkeys class option if keyword
                              %display desired
\maketitle

%\tableofcontents

A gas of bosons, when cooled down to a sufficiently low temperature, undergoes a phase transition forming a Bose-Einstein condensate (BEC) - a quantum degenerate state in which spontaneous coherence develops between the particles \cite{ketterle_bose_1995}. The general definition of BEC was proposed by Penrose and Onsager \cite{PenroseOnsager}. It is based on the single particle density matrix $\rho(x,x')=\langle \psi^\dag(x) \psi(x')  \rangle $ and states that BEC occurs when $N_c$, the largest eigenvalue of $\rho$, is of the order of the total number of particles in the quantum fluid. This definition remains meaningful for inhomogeneous systems and it corresponds to the existence of off-diagonal long range order (ODLRO), $\rho(x,x') \rightarrow n_c \neq 0$ for $|x-x'| \rightarrow \infty$ \cite{yang}. In practice however, the measurement of the single particle density matrix relies on challenging experiments. With ultracold atomic gases, the ODLRO was experimentally verified by a matter wave interference experiment \cite{Bloch2000}. To the best of our knowledge however, a full experimental reconstruction of the density matrix has not been performed yet. In this work, we used the favorable properties of microcavity polaritons to perform this task.

Polaritons are quasi-particles that represent the eigenmodes of the strong coupling regime between light and matter. Such a regime can be achieved in planar semiconductor microcavities~\cite{weisbuch_polariton_1992}, multilayered structures where a set of quantum-wells are placed at the antinodes of the cavity mode. This configuration allows for efficient coupling of the photon mode with the exciton resonance, yielding polariton eigenmodes for the strongly coupled system. Polaritons have a bosonic character and favorable properties for Bose-Einstein condensation: an extremely light effective mass coming from the photon component and interactions provided by the exciton component. So far, polariton condensation has already been demonstrated by several groups~\cite{kasprzak_bose-einstein_2006,balili_bose-einstein_2007,deng_spatial_2007,wertz_spontaneous_2010}.

Due to the finite transmissivity of the cavity mirrors, the polaritons have a life time, that is in the picosecond range. To maintain the polariton density, the losses are compensated by a continuous injection of excitons in the system. At some threshold injection intensity, the polariton phase space density becomes of order unity and stimulated emission sets in. This leads to a fast increase in the polariton density with excitation power and simultaneously, the onset of ODLRO is observed \cite{kasprzak_bose-einstein_2006}. While the life time of the polaritons is too short for the system to reach a global thermal equilibrium, a thermal distribution on the lower polariton branch was experimentally observed \cite{kasprzak_bose-einstein_2006}.

Thanks to their photonic component, the experimental situation for measurements of the single particle density matrix is significantly better for the BEC of exciton-polaritons. In fact, coherence measurements of polariton condensates are much easier than the ones with atomic gases. They can be done by optical interference experiments with the light that is emitted by the microcavity. The first unambiguous proof of Bose-Einstein condensation of microcavity polaritons was precisely obtained by such an experiment \cite{kasprzak_bose-einstein_2006}. The long-range spatial coherence was determined by interfering the light emitted by the microcavity with its inverted image, giving access to $\rho(x,-x)$.

In this Letter we report on the experimental reconstruction, through a series of optical interferometric experiments, of the full single particle density matrix for a one-dimensional polariton condensate, i.e. $\rho(x,x')$ for all $x$ and $x'$. The main motivation for the use of a 1D polariton state is the significantly smaller size of the density matrix with respect to the 2D case. The quantum tomography we perform on such a state allows us to test the Penrose-Onsager criterion for BEC in the framework of Bose-degenerate gas of polariton quasi-particles. We provide convincing evidence of the pertinence and validity of the Penrose-Onsager criterion even in the case of non-equilibrium BEC, as it is the case for the polariton system.

The sample is the same CdTe planar semiconductor microcavity of our previous works~\cite{kasprzak_bose-einstein_2006}. The sample features a pronounced disorder potential, naturally arising as a result of the epitaxial growth process: particular linear disorder landscapes have already been proved to allow for single-energy 1D polariton condensates~\cite{manni_cond_2011}. The quantum tomography - the density matrix reconstruction - requires the determination of both the amplitude and the phase of the first-order coherence function $g^{(1)}(x,d)\equiv \rho(x,x+d)$ of the condensate, where $x$ represents the spatial coordinate along the 1D condensate direction and $d$ is a given spatial separation between points of the condensate. Along with the coherence measurements, the condensate order parameter itself requires a characterization in terms of its density and phase structure. In order to measure the required quantities we make use of advanced interferometric techniques.

\begin{figure}[tb]
\includegraphics[width=0.45\textwidth]{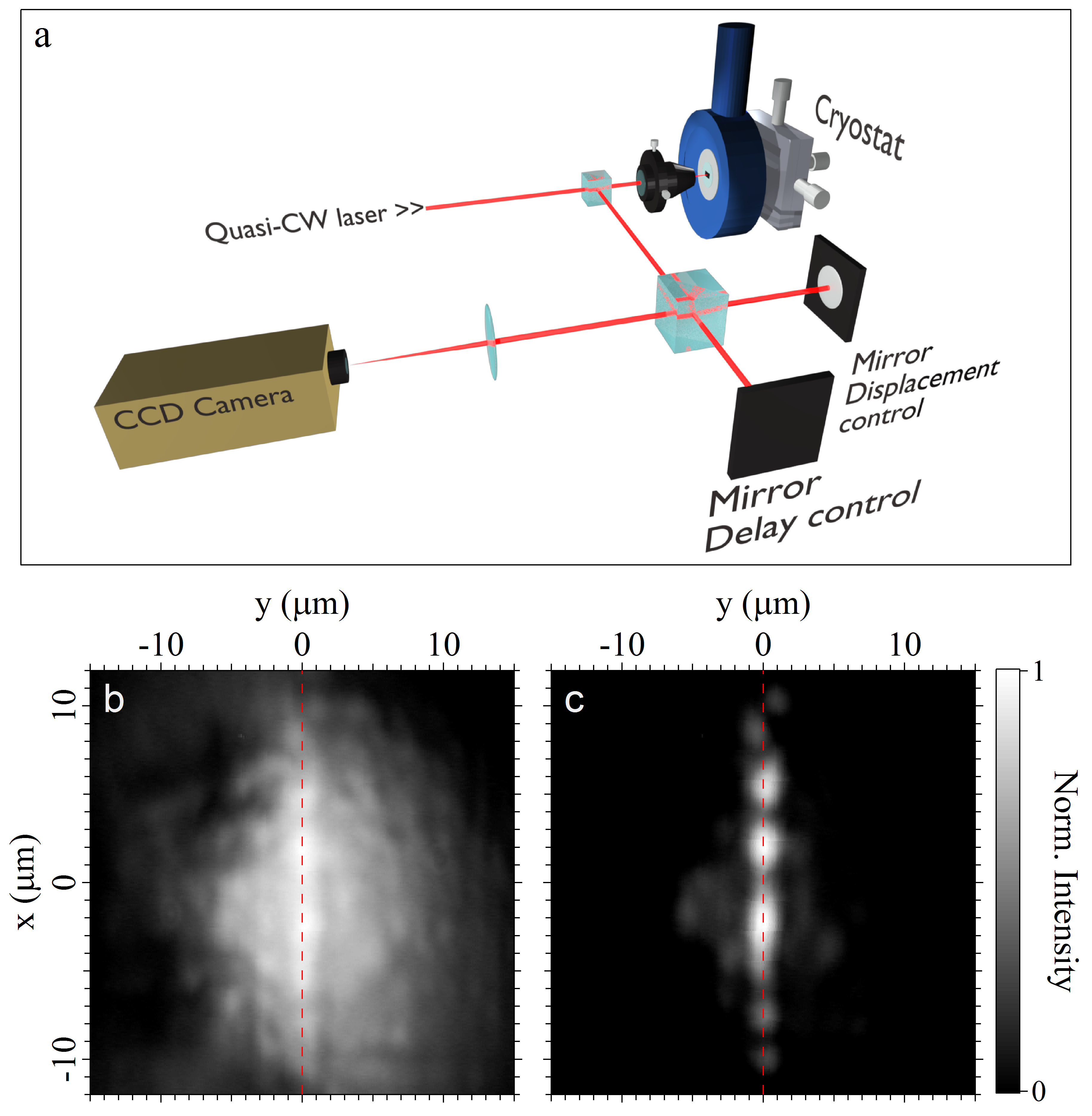}% Here is how to import EPS art
\caption{\label{fig:figure1} (a) Experimental setup. The non-resonant excitation laser is focused through a 0.5 NA microscope objective on the sample (hold in the cryostat at $\approx 4 K$). The same objectives collects the PL signal emitted by the sample, which is then sent at the input to the Michelson interferometer in the mirror-mirror configuration. One mirror is actively stabilized and controls the delay between the two arms; the other mirror is adjusted to realize the desired displacement $d$ between the two arms. Polariton density below (b) and above (c) condensation threshold. Condensation is observed to occur in a 1D state.}
\end{figure}

First we proceed to the determination of the first order coherence function. A schematic depiction of the experimental setup used for this purpose is shown in Fig.~\ref{fig:figure1}. The setup consists of a photoluminescence experimental apparatus in reflection configuration. The sample is held in a He-cooled optical cryostat at approximately 4 K. We shine a non-resonant Ti:Sapphire quasi-CW laser through a 0.5 NA microscope objective on the surface of the sample to create a polariton population in the system. The non-resonant excitation scheme ensures that no coherence is imprinted in the system by the laser onto the polariton gas. The PL signal emitted by the sample is collected through the same microscope objective and is send to an actively-stabilized Michelson interferometer~\cite{kasprzak_bose-einstein_2006}, which represents the core of the detection scheme and allows for high-stability phase measurements required for the purpose of this work. The interferometer, differently from our previous work on 1D condensates~\cite{manni_cond_2011}, is in the mirror-mirror configuration that is most convenient for the purposes of quantum tomography, as detailed in the following. The real-space condensate density is imaged by using a CCD camera placed at the output of the interferometer.

Starting from low excitation power, we observe emission from a classical gas of polaritons. By increasing the pump power, once we cross the threshold power of 35~$\mu W/cm^{2}$, we observe the spontaneous onset of condensation, characterized by the emission of a macroscopically populated single-energy polariton state. By slightly tilting one mirror of the interferometer (displacement control mirror in Fig.~\ref{fig:figure1}) we are able to set the overlap conditions between the two arms with micrometer resolution. Once an overlap condition is set, corresponding to a specific value of the $d$ parameter in the density matrix, a scan over $6\pi$ of the delay between the two arms of the interferometer is performed by shifting the other mirror of the interferometer (delay control mirror in Fig.~\ref{fig:figure1}). The scan of the delay is computer controlled by taking advantage of the active stabilization mechanism of the interferometer \cite{kasprzak_bose-einstein_2006}. This experimental procedure yields a discrete set of interferometric images, one for each delay condition of the interferometer arms. Thus, for each spatial point of the condensate, one obtains an intensity modulation as a function of delay. By fitting such intensity modulation we retrieve a map of the amplitude of the $g^{(1)}(x,d)$.

We consider a finite set of values for the displacement $d$, corresponding to steps of $\approx 1.2$ $\mum$, covering the whole length of the 1D condensate. The measurements are repeated for different excitation powers, below and above threshold. In particular, above threshold, we made sure to have only a single-energy condensed state emitting and to be far from the onset of multimode condensation.

As previously mentioned, access to the phase information is also required in order to fully reconstruct the density matrix. In order to experimentally assess this quantity, under the same excitation conditions, we make use of a different detection scheme. Such scheme basically consists of a Mach-Zehnder interferometer in which one arm is the polariton condensate density and the other arm is a magnified version of it. A small part of the 1D condensate, enlarged by four times, acts as an approximately flat phase reference and is overlapped with the whole condensate density, similarly to what was done in Ref.~\cite{manni_spontaneous_2011}. From the resulting interferogram, the spatial variation of the phase of the polariton field $\Delta \phi(x) = \phi(x)-\phi(x_0)$ is extracted. The density matrix is then constructed from the absolute value and phase variation as $\rho(x,x')=|\rho(x,x')| e^{i [\Delta \phi(x')- \Delta \phi(x)]}$. The fact that a non-constant phase exists for the single particle density matrix is due to the nonequilibrium state of the polariton quantum fluid, that breaks time invariance. Consequently, in contrast to equilibrium condensates \cite{stringari_bec}, the steady state does not correspond to the ground state of the system and its wave function is not guaranteed to be real. Within the polariton condensate, particles flow proportional to the gradient of the phase are allowed. This peculiarity of the nonequilibrium polariton condensates has made it possible to observe quantized vortices without setting the system into rotation \cite{lagoudakis_quantized_2008,lagoudakis_probing_2011}.

\begin{figure}[tb]
\includegraphics[width=0.5\textwidth]{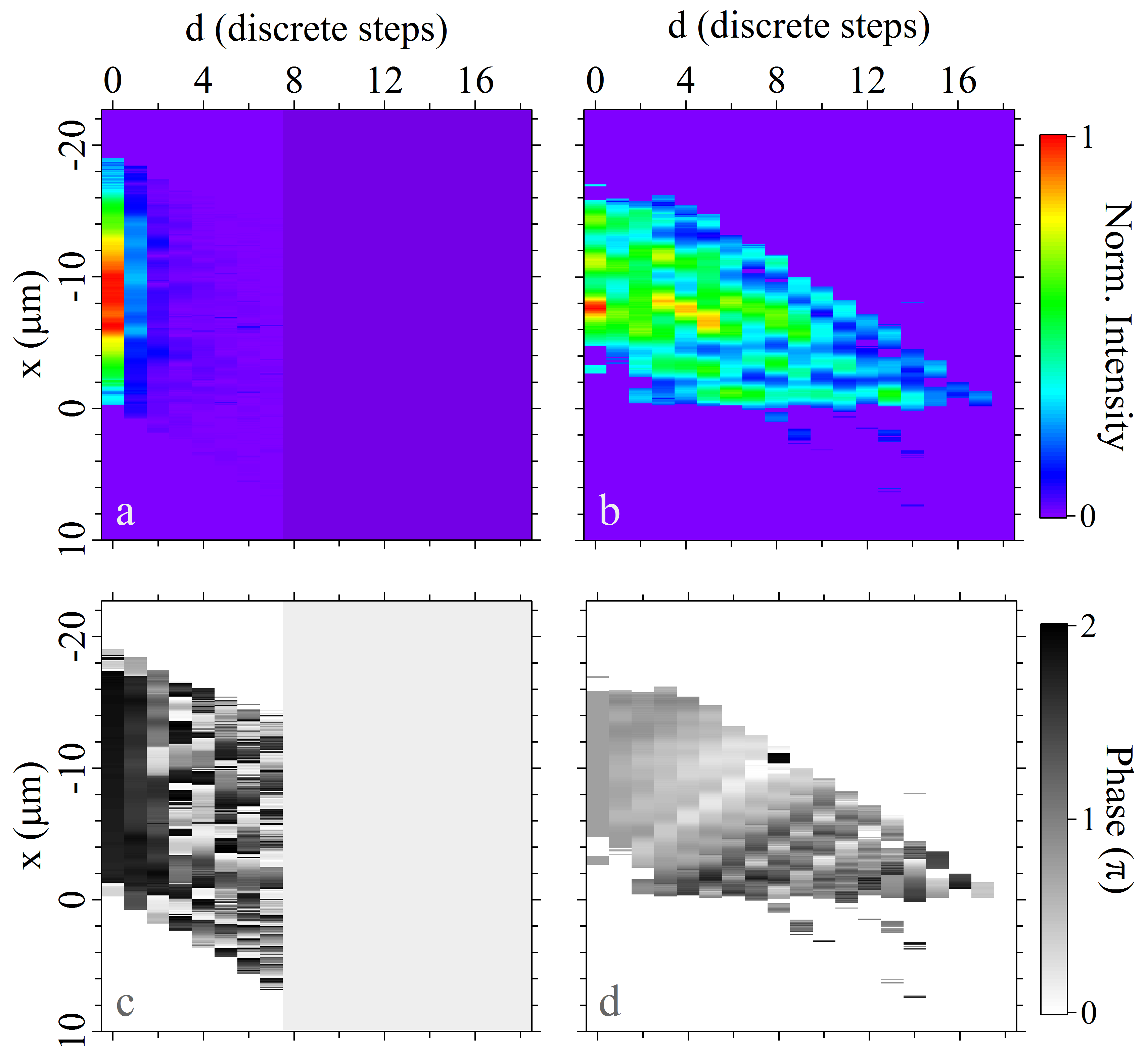}% Here is how to import EPS art
\caption{\label{fig:figure2} Density matrix reconstruction for the polariton system: amplitude of the $g^{(1)}(x,d)$ in the case below (a) and above (b) condensation threshold, respectively for pump powers of 20~$\mu W/cm^{2}$ and 60~$\mu W/cm^{2}$. The corresponding phase structure is shown in (c) and (d). A high degree of long-range order is present in the condensed phase as well as a defined phase structure. Below threshold the measured data stop at step $d = 7$.}
\label{figure2}
\end{figure}

The density matrices obtained through quantum tomography are shown in Fig.~\ref{fig:figure2}. Each column of the matrices represents the value of coherence along the condensate, between spatial points that are separated by a distance $d^{\prime} \approx d\cdot1.2$~$\mu m$ ($d = 0$ corresponds to full overlap whilst for $d = 18$ the two arms have a negligible overlap region). The amplitude of the $g^{(1)}(x,d)$, for a pump power below (20~$\mu W/cm^{2}$) and above (60~$\mu W/cm^{2}$) condensation threshold, is shown in Fig.~\ref{fig:figure2}(a) and (b) respectively. The difference between the two situations is striking: in the classical gas the coherence drops to zero within a few microns of displacement and becomes completely negligible already at step $d = 4$; in the condensate the off-diagonal long-range order (ODLRO) is well established and the coherence has significant values over the whole length of the condensate, even up to step $d = 17$. In Fig.~\ref{fig:figure2}(c) and (d), below and above condensation threshold, the phase of the density matrix is shown. Below threshold the relative phase is not well defined and it is seen to randomly fluctuate already for very small displacement values. On the contrary, above threshold a very well defined phase structure is identified, with a phase profile that varies smoothly as a function of spatial position $x$ and displacement $d$.

\begin{figure}[tb]
\includegraphics[width=0.5\textwidth]{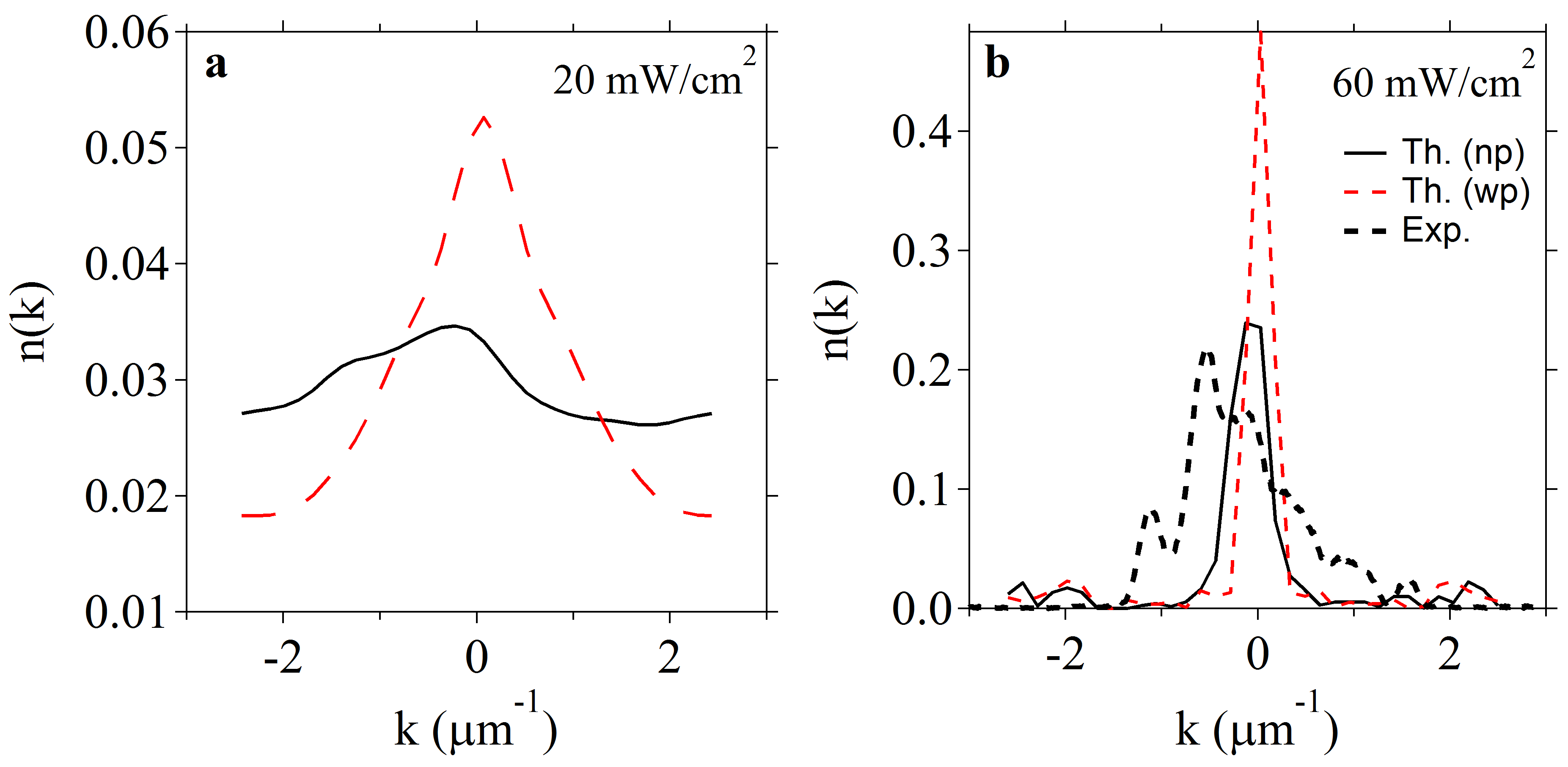}% Here is how to import EPS art
\caption{\label{fig:nk} The momentum distribution computed from the single particle density matrix containing the phase information (full line) and from $|rho|$ (dashed line), below (a) and above (b) threshold. When the phase information is omitted [Th. (np)], the momentum distribution is incorrectly centered around zero. With phase information [Th. (wp)], the distribution peak is shifted towards the left, towards the same direction as the measured k-space distribution (thick dashed line).}
\label{figure3}
\end{figure}

The correct phase of the density matrix is important to obtain the correspondence between the Fourier transform of the density matrix and the momentum distribution of the polariton gas.
\begin{equation}
n(k) = \int dx dx' \; e^{ik(x-x')} \; \rho(x,x').
\label{eq:nk}
\end{equation}
The momentum distribution reconstructed from the density matrix is shown in Fig.~\ref{fig:nk}. When the phase information of the single particle density matrix is retained, the momentum distribution is correctly peaked at a negative momentum. Instead, when the phase information is not included and a constant phase is assumed all over the condensate, the calculated momentum distribution appears centered around zero. The shift of the peak of the momentum distribution toward negative momenta corresponds to the experimentally measured momentum distribution and it is consistent with the measured phase gradient along the one-dimensional condensate that corresponds to a flow of polaritons with a finite wavevector. Comparison with the directly measured momentum distribution (see Fig.~\ref{fig:nk}(b), dashed thick line) however shows that the reconstruction of the momentum distribution from the density matrix is far from perfect: the experimental momentum distribution shows much more structure than the reconstructed one and its overall shape is much broader. We suspect that this is due to the finite spatial resolution in the phase measurement. This results in an underestimation of the spatial phase variations and consequently a momentum distribution that is too narrow.

\begin{figure}[tb]
\includegraphics[width=0.5\textwidth]{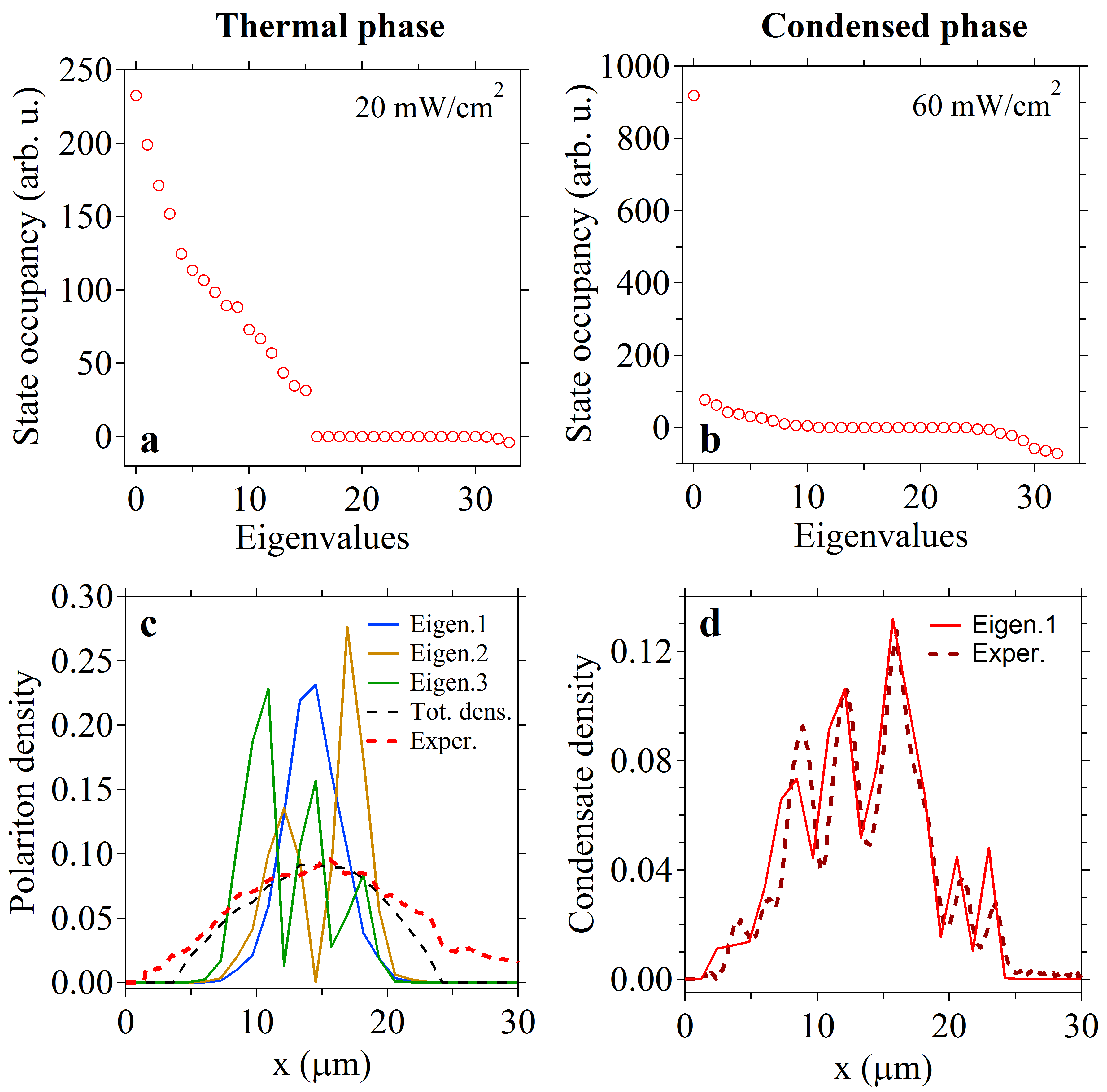}% Here is how to import EPS art
\caption{\label{fig:figure4} Results from the diagonalization of the density matrix. Below threshold (a,c): eigenvalues (a) and the eigenvectors (c) of the three states with the highest occupation. Above threshold (b,d): the eigenvalues (b) and the state of the condensate (d).}
\end{figure}

The diagonalization of the measured density matrix yields a set of eigenvalues, representing the population of the corresponding eigenstates for the system. The calculation of the eigenvalues allows us to directly validate the Penrose-Onsager criterion for BEC. For a classical gas, the quasi-particles are distributed among the different states available in the system, corresponding to a set of comparable non-zero eigenvalues. This is indeed what we find from the experimental data. Fig.~\ref{fig:figure4}(a) shows the occupations of the single particle states. As expected it has the shape of the Maxwell-Boltzmann distribution below threshold, corresponding to a Gaussian correlation function in real-space, as shown in Fig.~\ref{fig:figure4}(c). The densities of the three most occupied states are also plotted in Fig.~\ref{fig:figure4}(c). They show the spatial dependence of standing waves with increasing number of nodes.

On the contrary, upon polariton condensation in a single-energy state, only one state is macroscopically populated so that one eigenvalue should be found significantly different from zero and dominant over all the others [Fig.~\ref{fig:figure4}(b)]. This implies that a single-energy condensate was created. Co-existence of multiple condensates on the other hand would lead to several large eigenvalues \cite{leggett_bose_2001,leggett}. The position on the sample was actually carefully chosen for a single condensed state to occur. When scanning the position of the excitation spot, a place was chosen where a single line was dominant in the frequency spectrum of the condensate. The present analysis of the eigenvalues of the density matrix shows that this corresponds to a single-energy condensate. Indeed, a very good agreement between the macroscopically populated eigenstate and the measured polariton emission is found, as shown in in Fig.~\ref{fig:figure4}(d).

Unfortunately, we find non-physical negative occupation for the last few eigenvalues of the density matrix spectrum [see Fig.~\ref{fig:figure4}(d)]. We understand this spurious effect as a result of the experimental error on the determination of the coherence: such error is too large to attribute a physical meaning to the states that correspond to the weakly occupied states. The contribution of the condensate to the density matrix $\rho_c=N_c \phi^*(x_1) \phi(x_2)$ is so large that the experimental error on the remainder $\rho-\rho_c$ is too large to yield reliable results.

In conclusion, in this work we have performed a full measurement of the single particle density matrix of a nonequilibrium polariton condensate and we have completely determined the complex order parameter. The consistency of the Penrose-Onsager criterion for BEC above the threshold laser power was experimentally verified. This results give rigorous support and strengthens the analogy with the cold atoms BEC, reaffirming once more the central role of polaritons in the study of condensates and related phenomenology.

This work was supported by the Swiss National Science Foundation through NCCR ``Quantum Photonics'' and SNSF $\alpha$.project 135003. M.W. was supported by the UA.LP and FWO Odysseus programs.

\providecommand{\noopsort}[1]{}\providecommand{\singleletter}[1]{#1}%

\end{document}